\documentclass[9pt,twocolumn,twoside]{opticajnl}
\journal{opticajournal} % use for journal or Optica Open submissions
\setboolean{shortarticle}{false}
\usepackage{lineno}
\usepackage{amsmath} % amssymb
\usepackage{bbold}
\usepackage{graphicx}
\title{Jaynes-Cummings atoms coupled to a structured environment: Leakage elimination operators and the Petz recovery maps}

\author[1,*]{Da-Wei Luo}
\author[1,+]{Ting Yu}

\affil[1]{Center for Quantum Science and Engineering and Department of Physics, Stevens Institute of Technology, Hoboken, New Jersey 07030, USA}

\affil[+]{ting.yu@stevens.edu}
\affil[*]{dawei.luo@stevens.edu}

\begin{abstract}
	We consider the Jaynes-Cummings (JC) model embedded in a structured environment, where the atom inside an optical cavity will be affected by a hierarchical environment consisting of the cavity and its environment. We propose several effective strategies to control and suppress the decoherence effects to protect the quantum coherence of the JC atom. We study the non-perturbative control of the system dynamics by means of the leakage elimination operators. We also investigate a full quantum state reversal scheme by engineering the system and its coupling to the bath via the Petz recovery map. Our findings conclude that, with the Petz recovery map, the dynamics of the JC atom can be fully recovered regardless of Markov or non-Markovian noises. Finally, we show that our quantum control and recovery methods are effective at protecting different aspects of the system coherence.
\end{abstract}

\setboolean{displaycopyright}{false}

\begin{document}

\maketitle

\section{Introduction}

The Jaynes-Cummings model~\cite{Jaynes1963a, Shore1993a, Shore2004a} describes the interaction of a two-level atom with a quantized electromagnetic mode, and is a one of the most fundamental models in quantum optics~\cite{Allen2012a,Scully1997a, Agarwal2012a}. It is an exactly solvable model consisting of $2 \times 2$ diagonal blocks in the `bare' atom-field basis, supporting an exact analytical study while being capable of displaying a rich variety of physical phenomena, such as vacuum Rabi splittings~\cite{Zhu1990a}, light-matter entanglement~\cite{Phoenix1991a}, and a periodic collapse and revival of quantum coherence~\cite{Eberly1980a}. The JC model has also been generalized~\cite{Larson2021a} to three~\cite{Yoo1985a} level atoms, $N$ two level atoms~\cite{Tavis1968a} (which has been coined the `Tavis-Cummings model'), and the quantum Rabi model~\cite{Rabi1936a,Braak2011a} which also includes the terms dropped by the rotating wave approximations. The entanglement between multiple JC atoms~\cite{Yu2004b, Yu2009a, Yonac2006a} can be found to show a feature of the sudden death and birth of quantum entanglement. Analytical methods to study the JC model has also triggered the development of analytical tools that can be applied to other interesting models~\cite{Shore1993a, Larson2021a}, such as the algebraic approach~\cite{Vadeiko2003a, Bonatsos1993a} with deformed algebras to solve for the eigenvalue problem of the Tavis-Cummings model or other extensions of the JC model such as with intensity dependent couplings~\cite{Singh1982a}. Experimental realizations of the JC model has also been studied extensively~\cite{Shore1993a}, most notably with the use of Rydberg atoms and one-atom masers~\cite{Diedrich1988a,Rempe1987a,Meschede1985a,Brune1996a}.

Quantum systems such as JC atoms and the optical cavity inevitably interact with their surroundings, so it is desirable to study JC systems in an open quantum system framework~\cite{Breuer2002}. With the rapid pace of utilizing quantum technologies, one of the most pressing needs is the understanding of the open system dynamics, as well as ways to control or protect the system in a generic open system setting. For the study of open quantum system dynamics, while the Markov approximation that discards the memory effects of the quantum bath can be a valid choice in the case of weak coupling with an unstructured bath~\cite{Breuer2002}, many important situations such as high-Q cavity, strong coupling regimes, and many realistic applications will need to consider the memory effects and non-Markovian dynamics. To this end, various theoretical and mathematical tools~\cite{Strunz1999, Diosi1998, Bellomo2007, Breuer2002} have been developed such as the influence path integral~\cite{McKane1990a, Hanggi1989a} and non-Markovian quantum jumps~\cite{Piilo2008}. Notable among them is the quantum state diffusion (QSD) equation approach, which uses stochastic quantum trajectories and their ensemble average to capture the open system dynamics~\cite{Diosi1998, Strunz1999}. The QSD approaches has been applied to the exact solution of many interesting physical models such as Brownian motions~\cite{Strunz2004a} and multi-level atoms~\cite{Jing2012, Jing2010a}, can be approximated perturbatively~\cite{Yu1999a} or numerically solved with functional derivative expansions~\cite{Luo2015a}. The resource for most quantum-enhanced technologies such as entanglement~\cite{Yu2009a, Yonac2006a, Vedral2014a, Horodecki2009a} and quantum coherence~\cite{Streltsov2017a} can be susceptible to such environmental effects, necessitating some strategy to mitigate or co-exist with the detrimental effects of the open system dynamics. To date, various methods have been proposed, such as the dynamical decoupling pulses~\cite{Viola1998a,Viola1999a}, bath engineering~\cite{Murch2012a, Poyatos1996a}, and decoherence free subspaces~\cite{Beige2000a, Lidar1998a}. Here, we consider two different approaches to control the dissipation for a leaky JC model. The first is based on the so-called leakage elimination operators (LEO)~\cite{Jing2015a, Zheng2020a}, which uses a projection to divide the Hilbert space and derive a control to suppress the leakage between the subspaces. The second is to employ the Petz recovery maps~\cite{Petz1986a, Petz1988a}, first introduced in the context of subalgebras and relative entropy. The Petz recovery map has been applied to quantum state  discrimination~\cite{Hausladen1994a,Gilyen2022a,Mishra2023a}. In addition, its ability to achieve a near-optimal reversal quantum dynamics is studied in~\cite{Barnum2002a}. Very recently, a discrete version has been proposed~\cite{Gilyen2022a}. It has then been applied to Markov open system dynamics~\cite{Alhambra2017a}, and has been cast into a Lindblad master equation form~\cite{Kwon2022a} with the ability to fully reverse the dynamics of a single quantum trajectory, and construct a protected code space to use in conjuncture with error correction codes. While there are discussions on the possibility of approximated recover or reversal of dynamics~\cite{Taranto2021a,Lautenbacher2022a} in the context of non-Markovian effects, a dynamical approach to reverse a non-Markovian quantum trajectory has not yet been studied in detail.

In this paper, we will consider the system of a lossy JC model embedded in a non-Markovian bosonic bath, and study the control of open system dissipation for the JC system. This model can, for the two-level atom inside the optical cavity, constitute a hierarchical structured bath: one direct bath (i.e., the optical cavity) and an indirect bath (the bosonic environment). We will first consider how to construct an appropriate LEO to suppress the environmental noise on the system, and study the effectiveness of various control fields. This approach represents the paradigm of applying a control on the quantum system alone to isolate it from the influences of its environment. Furthermore, we will use the Petz recovery map to study how to reverse the non-Markovian open system dynamics for the JC atom. Finally, we discuss how this process relates to the non-Markovianity of the JC atom dynamics. The two decoherence control protocols considered here represents different paradigms: the LEO suppresses the leakage and is applied as the system is evolving, while the Petz map based control is applied after the system has evolved dissipatively and drives the state back along the forward trajectory.

\section{Decoherence control with leakage elimination operators}

Consider the Jaynes-Cummings model embedded in a bosonic bath,
\begin{align}
	H_{\rm tot} &= \frac{\omega}{2}\sigma_z+\kappa[\sigma^-a ^\dagger+\sigma^+ a] + \omega_c a ^\dagger a \nonumber \\
	            & + \sum_k g_k [a b_k ^\dagger + a ^\dagger b_k] + \sum_k \omega_k b_k ^\dagger b_k \nonumber \\
	            & \equiv H_s + H_{\rm int} + H_b,
\end{align}
where $\sigma_z$ is the Pauli-$z$ operator for the two-level atom, with the frequency of $\omega$. $\sigma^{+(-)}$ are the atom's raising/lowering operators, and $a(a ^\dagger)$ the annihilation (creation) operators for the single-mode cavity with the frequency of $\omega_c$, and $b_k(b_k ^\dagger)$ the annihilation (creation) operators for the $k$th environmental mode with the frequency $\omega_k$. Here $\kappa$ denotes the atom-cavity coupling strength, and $g_k$ represents the coupling strength between the cavity and the $k$th bath mode. The total Hamiltonian includes the system ($H_s$), bath ($H_b$) and system-bath interaction ($H_{\rm int}$) parts, where the interaction part may be written as a Lindblad operator interacting with the bath, $L\sum_k g_k b_k^\dagger + h.c.$, where $L=\lambda a$ with $\lambda$ being the system-bath interaction strength. In this model, the JC atom will experience a structured hierarchical environment consisting of a cavity mode and the bosonic bath (indirectly).  The influences of the bath may be encoded in a bath correlation function $\alpha(t, s) = \int d\omega J(\omega)e^{-i\omega(t-s)}$, where $J(\omega)$ is known as the bath spectrum $J(\omega) = \sum_m |g_m|^2 \delta(\omega - \omega_m)$. Here, we will be focusing on the Lorentzian spectrum
\[
	J(\omega) = \frac{1}{2\pi} \cdot \frac{\gamma^2}{(\omega-\omega_0)^2+\gamma^2} \, ,
\]
corresponding to a memory function of the form $\alpha(t,s) = \frac{\gamma}{2}\exp(-i\omega_0(t-s) - \gamma|t-s|)$. Here the parameter $\omega_0$ is a central frequency shift, and $\gamma$ signifies the memory time or strength: smaller $\gamma$ means stronger memory effects (longer memory time $1/\gamma$), and $\gamma \rightarrow \infty$ is the memory-less Markov limit. This bath spectrum allows us to tune the memory strength of the bath in a continuous fashion to display the Markov-nonMarkovian transition features. This generic open system treatment can capture some intricate dynamical behaviors that have been seen in the Markov cases. Here we have assumed $\gamma_{\rm eff} = \gamma + i\omega_0$, the model is simple enough to allow some analytical analysis, yet presents very rich physics. For example, it has been show that~\cite{Ma2014g} by tuning the system parameters, the dynamics of the JC atom may be non-Markovian even if the bath is memory-less (Markov), and vice versa. The term ``non-Markovian'' or ``Markov'' for the dynamics here would refer to whether the dynamics of the JC atom shows memory effects, signified by a non-monotonic trace distance change~\cite{Breuer2012a,Rivas2014b}, or a flow of information from the environment to the system~\cite{Lu2010a, Chrui2018a, Zhong2013a}.

Since quantum systems inevitably interact with their surroundings and lose quantum coherence in the process, various strategies have been proposed to combat such effects. Among many quantum control schemes, the dynamical decoupling pulses (`bang-bang' pulses)~\cite{Viola1998a, Viola1999a} have been proven successful at suppressing the detrimental effects resulting from the open system dynamics. The decoupling pulses are assumed to be very short-lived but large in amplitude, such that in the times $\Delta t$ when the pulse is active it is dominating and the system dynamics is ignored during the pulse active time. As a perturbation-free alternative, the leakage elimination operators (LEO) have been proposed where the decoupling control is treated exactly alongside with the system dynamics using a PQ-partition projection technique. By projecting the Hilbert space into a projective $\mathcal{P}$ space and its complement $\mathcal{Q}$ space, with the projector $P^2=P, Q^2=Q, Q = I-P$. Mathematically, we may denote the projector $P=\sum | \nu_i \rangle\langle \nu_i|$ where the sum is over the orthonormal basis for the projective space $|\nu_i \rangle$. The Hamiltonian and quantum state can also be partitioned as $H_P = PHP$, $H_Q = QHQ$ and the transition or `leakage' between the $\mathcal{P}$ and $\mathcal{Q}$ spaces $H_L=PHQ+QHP$. It has been shown that an operator $R$ can suppress the leakage if it acts as identity in $\mathcal{P}$, minus identity in $\mathcal{Q}$ and anti-commutes with the leak, i.e. $[R,P]=[R,Q]=\{R, H_L\}=0$, $R=\mathrm{diag}(c_1 I_P, -c_2 I_Q)$, where $c_{1,2}$ are control functions. Here, we choose $c_1=c_2=c(t)$, and the system Hamiltonian with the LEO control is denoted as
\begin{equation}
	\widetilde{H}_s=H_s+c(t)R.
\end{equation}
The LEO can effectively cancel out the leakage between the \(\mathcal{P}\) and \(\mathcal{Q}\) space when the control $c_{1,2}$ satisfy the pulse requirements~\cite{Jing2015a} by inducing large oscillations inside the integral associated with the leak such that the integral would cancel out over time. The resulting control can be forgiving in that only the integral of the control fields plays a critical role, and fluctuations or errors in the control can be tolerated.

The non-Markovian open system dynamics of the system can be exactly and analytically solved. To date, many different approaches to deal with the non-Markovian dynamics has been developed~\cite{Bellomo2007,Piilo2008,Diosi1998,Strunz1999, Breuer2002}. Here, we use the Quantum State Diffusion (QSD) equations~\cite{Strunz1999, Diosi1998, Yu1999a} to study the dynamics. By projecting the bath modes onto a Bargmann coherent state labelled by $\langle z^*_t|$, a Schr\"odinger type equation can be written for each quantum trajectory $|\psi_t(z^*) \rangle$,
\begin{align}
	\partial_t |\psi_t(z^*_t) \rangle = \left[-i\widetilde{H}_s + Lz_t^* - L ^\dagger \bar{O}(t,z^*)\right]|\psi_t(z^*_t) \rangle,
\end{align}
where we introduce a noise $z^*_t=-i\sum_k g_k z^*_k e^{i \omega_k t}$, with $z_k$ labelling the coherent state of the $k$th mode, and the $\bar{O}$ operator denotes an ansatz for the functional derivative
\begin{align}
	& O(t,s,z^*) |\psi_t(z^*_t) \rangle = \frac{\delta}{\delta z_s^*} |\psi_t(z^*_t) \rangle, \nonumber \\
	& \bar{O}(t,z^*) = \int_0^tds \alpha(t,s)O(t,s,z^*).
\end{align}
The $\bar{O}$-operator follows a consistency condition
\[
	\frac{\partial}{\partial t}\frac{\delta}{\delta z_s^*} |\psi_t(z^*_t) \rangle = \frac{\delta}{\delta z_s^*} \frac{\partial}{\partial t} |\psi_t(z^*_t) \rangle,
\]
which leads to
\begin{align}
	\dot{O}(t,s,z^*) &= \left[ -i\widetilde{H}_s + Lz_t^* - L^\dagger \bar{O}(t,z^*), O(t,s,z^*) \right] \nonumber \\
					 & -L ^\dagger \frac{\delta \bar{O}(t,z^*)}{\delta z_s^*}
\end{align}
The $\bar{O}$ operator may then be obtained either exactly~\cite{Jing2012, Jing2010a, Strunz2004a} or with approximations~\cite{Yu1999a} or numerically~\cite{Suess2014a,Luo2015a}. The reduced density operator can be recovered by taking an ensemble average of the trajectories, $\rho_s = \mathcal{M}\left[|\psi_t(z^*_t) \rangle \langle \psi_t(z^*_t) |\right]$. Notably, a master equation can also be accordingly derived using the QSD using the Novikov's theorem~\cite{Yu1999a, Novikov1965a}, and in the case where the $\bar{O}$-operator is itself noise-free, the master equation takes the form (denote the noise-free $\bar{O}(t,z^*) \rightarrow \bar{O}(t)$),
\begin{align}
	\frac{\partial}{\partial t} \rho_s(t) &=-i \left[\widetilde{H}_s, \rho_s(t)\right] % \nonumber \\
	+\left[L,\rho_s(t)\bar{O}^\dagger(t)\right]-\left[L ^\dagger,\bar{O}(t)\rho_s(t)\right]. \label{eq_meq}
\end{align}

\begin{figure}
	\centering
	\includegraphics[width=.41\textwidth]{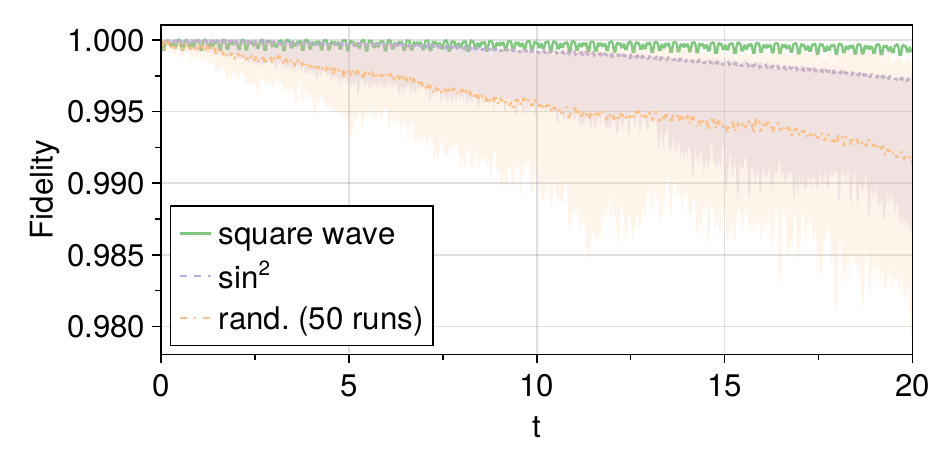}
	\caption{Using LEO, the fidelity dynamics between the evolved state and the ideal state evolving under the free Hamiltonian $H_p$, with ideal square wave (green solid line), ideal sine-squared wave (purple dashed line), and the average of 50 runs of square waves with random errors (orange dash dotted line). The shaded areas show the range of the 50 random trials for the sine-squared wave with errors (shaded purple) and square wave with errors (shaded orange).}\label{fig_leo_cft}
\end{figure}

Choosing the $\mathcal{P}$-space projective operator to be $P=|g0 \rangle \langle g0| + |e0 \rangle \langle e0 |$, it can be shown that $R=c(t)[|g0 \rangle \langle g0| + |e0 \rangle \langle e0 | - |g1 \rangle\langle g1|]$ can act as a LEO to protect the dynamics of the system in the $\mathcal{P}$-space for the TLS. The corresponding effective Hamiltonian is $H_P=\omega \sigma_z/2$, and an ideal LEO would allow an initial in the $\mathcal{P}$-space to evolve under a unitary propagator $U_p(t) = \mathcal{T}\exp(-i\int_0^t ds H_P)$. With an initial state $|\varphi_c(0) \rangle =  [|g0\rangle + |e0 \rangle]/\sqrt{2} \in \mathcal{P}$, we denote the ideal evolved state as $|\varphi_c(t) \rangle = U_p(t) |\varphi_c(0) \rangle$. Using the consistency condition, we can exactly solve for the $\bar{O}$-operator in this case. The $\bar{O}$ operator is found noise independent and is in the form of $O(t,s) = f_1(t,s) |g0 \rangle\langle e0| + f_2(t,s) |g0 \rangle\langle g1|$, where
\begin{align}
	\partial_t f_1(t,s) &= \left[i \kappa + \lambda F_1(t) \right] f_2(t,s) + i \omega f_1(t,s), \nonumber \\
	\partial_t f_2(t,s) &= i \left[\kappa f_1(t,s) + \left[\omega _c-2 c(t)\right] f_2(t,s)\right]
		+ \lambda f_2(t,s) F_2(t), \label{eq_leo_f12}
\end{align}
with boundary conditions $f_1(t,t)=0$, $f_2(t,t)=\lambda$, and $F_i(t) = \int_0^t ds \alpha(t,s) f_i(t,s)$. It can be seen that in this case, the $F_1$ term is not directly dependent on the control $c(t)$ but is influenced indirectly through $F_2$. To quantify the effectiveness of the dissipation control, we use the fidelity~\cite{Jozsa1994a,Nielsen2000a}
\begin{align}
	F(\rho, \sigma) = \left( \mathrm{Tr} \sqrt{\sqrt{\sigma} \rho \sqrt{\sigma}} \right)^2,
\end{align}
between the evolved state and the ideal state $|\varphi_c(t) \rangle$, choosing $\omega=\omega_c=1$, $\lambda=0.6$, $\kappa=0.7$, $\gamma_{\rm eff}=0.4$. The result is shown in Fig.~\ref{fig_leo_cft}. We first consider an equal-width periodic square wave $c(t) = A$ for $t\in [\tau_c/2, \tau_c]$ and $c(t)=0$ otherwise, for each period $\tau_c$. Taking $\tau_c=0.1$ and amplitude $A=100$, it can be seen the square wave works well to keeping the fidelity close to $1$. An ideal sine-square wave $A \sin^2(\omega_p t)$ with the same period ($\omega_p=10\pi$) is also considered, which can also suppress the leakage of the TLS. To study the robustness of the control fields, we also include $\pm 5\%$ random errors in the both the amplitudes and phases of the sine-square wave. For the square wave, the amplitudes and raising/falling edges are subjected to the errors, and the average and range of $50$ runs of the random errors are also shown in Fig.~\ref{fig_leo_cft}. It can be seen that they are quite resilient to the random noises, and can still maintain a high fidelity $>0.99$ on average. This shows that with the LEO approach, we have freedom in choosing various control fields and it can also be resilient against imperfections in the control as long as the control condition~\cite{Jing2015a} is met where the integral $\int ds c(s)$ is sufficiently large.

\begin{figure}
	\centering
	\includegraphics[width=.46\textwidth]{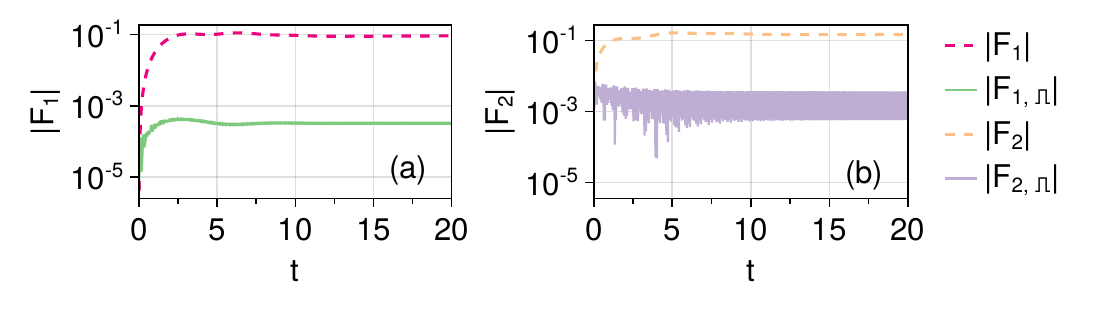}
	\caption{The norm of the $\bar{O}$ operator as a function of time, using the square wave as example. Panel (a) $|F_{1}|$ without (red dashed line) and with (green solid line) the control of LEO. Panel (b) $|F_{2}|$ without (orange dashed line) and with (purple solid line) the control of LEO.}\label{fig_leo_onorm}
\end{figure}

To get an insight into how the controls are effective in suppressing the leakage, we note that the solution to~\eqref{eq_leo_f12} can be formally written in the general form similar to~\cite{Jing2015a} as
\begin{align}
	F_2(t) \propto e^{-\gamma_{\rm eff} t}\int_0^t ds e^{-i \omega_c s + \gamma_{\rm eff}s} \exp\left[2i \int_0^s ds' c(s') \right] G(s), \label{eq_o_f2}
\end{align}
where
\begin{align}
	G(s) = \lambda \left[\frac{\gamma}{2} + F_2{}^2(s) \right] + i \kappa  F_1(s).
\end{align}

The mechanism behind the LEO is that when integral of the control $c(s')$ is sufficient large, it can lead to the integral of a comparatively-flat kernel $K(s)$ under a fast-oscillating envelope, formally as
\begin{equation}
	\int_0^t ds e^{i\int_0^s ds' c(s')} K(s),
\end{equation}
such that the kernel $K(s)$ may be considered constant in each period of the oscillating envelope $e^{i\int_0^s ds' c(s')}$ in the integral. Thus, the integral can cancel out in time when the frequency of the envelope is much larger than the typical evolution time of the kernel. Here, we can see numerically that the in the kernel function $G(s)$, the $F_2{}^2(s)$ term is around two orders of magnitude smaller than $F_1(s)$, and the leading order term in $G(s)$ is the constant $\lambda \gamma/2$, $F_1(s)$ is also slow-varying, which satisfies the LEO requirements. Note here that while $F_1$ does not directly depend on the control field $c(t)$, its solution follows
\begin{align}
	F_1(t) &\propto e^{-\gamma_{\rm eff} t} \int_0^t ds e^{-i \omega s + \gamma_{\rm eff} s} \exp\left[- \lambda \int_0^s ds' F_2(s') \right] F_2(s), \label{eq_o_f1}
\end{align}
whose suppression from the LEO is indirect via $F_2(t)$. In Fig.~\ref{fig_leo_onorm} we show the norms of the $\bar{O}$ operator's elements as a function of time. It can be seen that the LEO can suppress the norms of the $\bar{O}$ operator by about 2 orders of magnitude, taking the ideal square wave as an example. With smaller $|F_{1,2}(t)|$, the dissipative terms in the master equation~\eqref{eq_meq} is suppressed and we get the desired decoherence control.

% Alternatively, another viable LEO operator is $R'=\mathrm{diag}(-1, 1, 1)$ in the bare state basis $\{|g0 \rangle, |e0 \rangle, |g1 \rangle\}$, with a corresponding P-space $\mathrm{diag}(0,1,1)$. This P-space protects the coherence between the 1-excitation states $|e0 \rangle$ and $|g1\rangle$, but the coherence term of the TLS after tracing out the cavity mode is not covered. Note that this LEO is equivalent to the V-type LEO in~\cite{Jing2015a} for the three-level system, where the three levels correspond to the three dressed states in this subspace, and we would get the same result as referenced.

\section{Reversing non-Markovian trajectories with the Petz recovery maps}

The Petz recovery maps were first proposed in the context of quantum sufficient subalgebras~\cite{Petz1988a,Petz1986a}, and has been applied to measurement protocols to distinguish quantum states in an ensemble~\cite{Hausladen1994a,Gilyen2022a,Mishra2023a}. Due to its mathematical complexity, while its ability to reverse quantum dynamics has been suggested~\cite{Barnum2002a}, it is not until recently that an implementation of the Petz recovery map was suggested in the form of a master equation~\cite{Kwon2022a}. Approximated map using static reference states have also been studied in the context of quantum channels~\cite{Lautenbacher2022a}.

Consider a generic quantum channel $\mathcal{N}$ and an initial state $\rho_0$ as a reference state, the Petz recovery map can be given as a composite map of three separate maps
\begin{equation}
	\mathcal{R}_{\rho_0, \mathcal{N}}(\cdot) = \rho_0^{1/2}\; \mathcal{N}^\dagger\left[\sigma^{-1/2}(\cdot)\sigma^{-1/2}\right]\; \rho_0^{1/2} \label{eq_petzdef}
\end{equation}
where $\sigma=\mathcal{N}(\rho_0)$, and the action of the quantum channel $\mathcal{N}$ may be written using the Kraus operators~\cite{Kraus1983a} $N_k$ satisfying $\sum_k N_k ^\dagger N_k = I$ as $\mathcal{N}(\rho) = N_k \rho N_k^\dagger$, whose conjugate is defined as $\mathcal{N}^\dagger(\rho) = N_k^\dagger \rho N_k$.
The three individual maps $(\cdot) \rightarrow \sigma^{-1/2}(\cdot)\sigma^{-1/2}$, $(\cdot) \rightarrow \mathcal{N}^\dagger(\cdot)$ and $(\cdot) \rightarrow \rho_0^{1/2} (\cdot) \rho_0^{1/2}$ are positive maps but not necessarily trace-preserving themselves. However, the resulting composite map is positive and trace preserving on the support of $\sigma$~\cite{Kwon2022a, Gilyen2022a}. For a Markov master equation in the Lindblad form~\cite{Lindblad1976a, Breuer2002}
\begin{align}
	\partial_t \rho_t = \mathcal{L}(\rho_t) &= -i [H_s, \rho(t)] + \sum_n L_n \rho_t L_n ^\dagger - \frac{1}{2}\{L_n ^\dagger L_n , \rho_t\} \nonumber \\
	                                        &= -i [H_s, \rho(t)] + \sum_n \mathcal{D}_n[L_n]\left(\rho_t\right),\label{eq_mmeq}
\end{align}
one can decompose each infinitesimal time step and construct the recovery map for each time step and taking the continuous time limit to obtain~\cite{Kwon2022a} an explicit form of the reversal master for~\eqref{eq_mmeq} written in the same Lindblad form for the reversal time $t'=0 \rightarrow \tau$
\begin{align}
	\partial_{t'} \rho_{t'}^B &= \mathcal{L}_B(\rho^B_{t'}) \nonumber \\
	                          &= -i [H_r(\rho_{\tau - t'}), \rho_{t'}^B] %
	+ \sum_n \mathcal{D}_n[L_{r,n}(\rho_{\tau-t'})] \left(\rho_{t'}^B\right),\label{eq_mmeqR}
\end{align}
with reversal Hamiltonian $H_r$ and Lindblad operator $L_r$ given by
\begin{align}
	H_r(\rho_t) &= -H_s + \sum_{\eta {}_t, \eta {}_t', n}
	c_M(\eta_t, \eta'_t)
	\left\langle \eta_t \middle| M_n \middle| \eta_t' \right\rangle |\eta_t \rangle \langle \eta'_t| \nonumber \\
	                & \equiv -H_s + H_c, \nonumber \\
	L_{r,n}(\rho_t) &= \rho_t^{1/2} L_n ^\dagger \rho_t^{-1/2} \label{eq_hlrev}
\end{align}
where, denoting the eigen-decomposition of the density operator as $\rho_t = \sum_\eta \eta_t |\eta_t \rangle\langle \eta_t|$ for the composite atom-plus-cavity system,
\begin{align}
	 & c_M(\eta_t, \eta'_t) = -\frac{i}{2} \frac{\sqrt{\eta_t} - \sqrt{\eta'_t}}{\sqrt{\eta_t} + \sqrt{\eta_t'}}, %\nonumber \\
	\quad M_n = L_{r,n}^\dagger L_{r,n} + L_n ^\dagger L_n.
\end{align}

\begin{figure}[t]
	\centering
	\includegraphics[width=.4\textwidth]{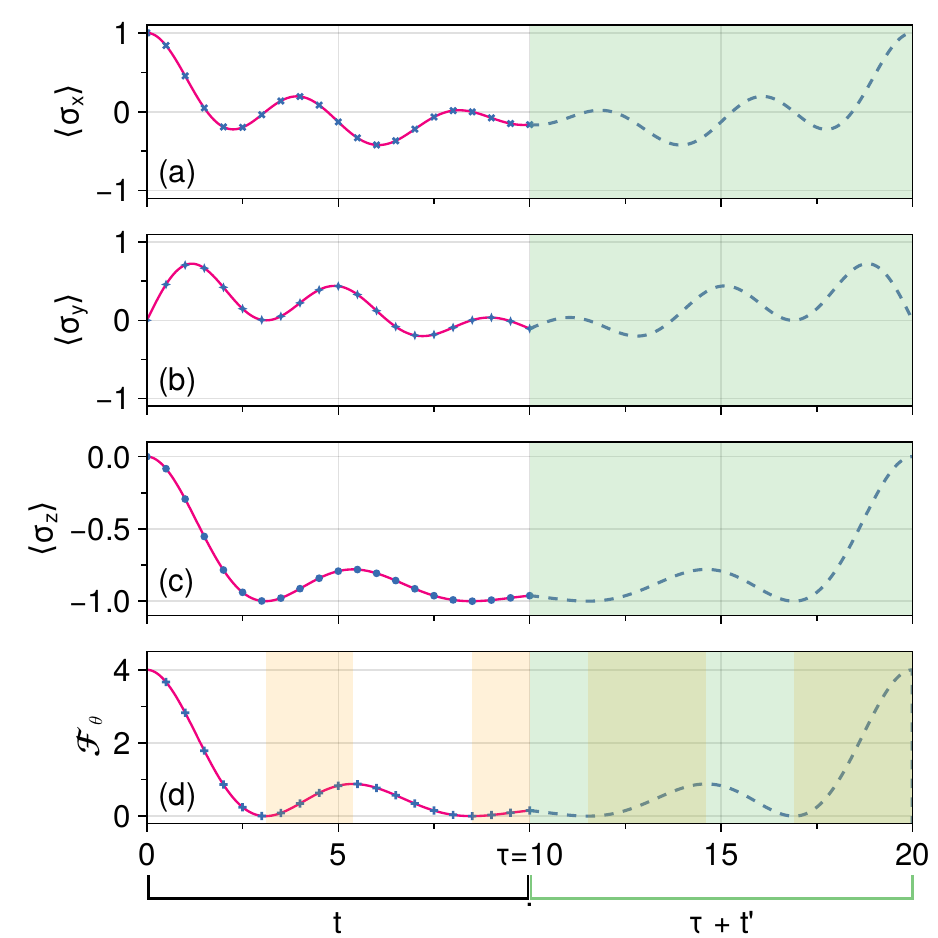}
	\caption{Non-Markovian dynamics of the atom in a hierarchical environment, and its reversal via the Petz recovery map. Shaded green area represents the time when the reversal control is being applied. Solid pink lines are the non-Markovian forward dynamics without external control, the dashed blue lines are under the Petz recovery map control, which is also plotted in-reverse for $\tau-t'$ with cross and circle markers to emphasize that the controlled dynamics is the exact mirror image for $\tau \rightarrow 2\tau$. Panel (a) - (c): expectation values for $\langle \sigma_{x,y,z} \rangle$, and Panel (d) shows the Fisher information $\mathcal{F}_\theta$ as a function of time. The orange-shaded bands are regions where the dynamical map is non-contractive, during which times the information can be observed to flow from the environment back into the system.}\label{fig_petz_rev}
\end{figure}

For the dissipative JC model under consideration, it has been shown that even if the bath is memory-less ($\gamma \rightarrow \infty$), the TLS atom can still have non-Markovian dynamics when the atom-cavity coupling is larger than $1/4$ of the effective coupling between the cavity and the bath~\cite{Ma2014g}. Therefore, in this parameter region, by reversing the Markov dynamics of the dissipative JC model, from the standpoint of the atom, it would be as if its non-Markovian dynamics is being reversed. %In Fig.~\ref{fig_petz_r_hlnormn} we show the norms of the reversal Hamiltonian and the reversal Lindblad operator,
As an example of using the reversal map, here we take $\omega_c=\omega=1$, $\kappa=0.6$, and $\lambda=0.75$, with an initial pure state $|\psi(0) \rangle = (|g0 \rangle + |e0 \rangle)/\sqrt{2}$, evolving for a duration of $\tau=10$. In Fig.~\ref{fig_petz_rev} we show the expectation values $\langle \sigma_{x,y,z} \rangle$ of the atom for the forward dissipative evolution for the duration of $[0,\tau]$, and the controlled reversal dynamics for $[\tau,2\tau]$ in panels (a) - (c). The backwards evolution time $t'=0 \rightarrow t$ begins after $\tau$. The reversal dynamics is both plotted as dashed lines for the time $\tau+t'=[\tau, 2 \tau]$, and shown in reverse for times in $\tau-t'=[0,\tau]$ to emphasize that the reversal dynamics backtracks the forward non-Markovian dynamics of the TLS.
It can be seen that the reversal dynamics are mirror images of the forward dynamics, and at the final state of the reversal dynamics recovers the initial state. Using the recovery master equation via the Petz recovery map, we can not only recover the initial state but also sending it back along the same path of the forward dissipative dynamics.

Intuitively, as the Petz recovery map is able to fully recover the initial state, one would expect that the information lost in the dissipative evolution is also being recovered after the control. Such back-flow of information suggests a non-Markovian evolution for the recovery process. To gain more insight, we study the non-Markovian features of the Petz recovery dynamics. There are various measures and criteria for the non-Markovian nature of open system dynamics~\cite{Breuer2012a, Vacchini2011a,Breuer2016a, Rivas2010a, Sutherland2018a, Fanchini2014u, Li2018a, Chrui2018a, Milz2019a}. One widely studied measure is constructed in terms of non-monotonic trace distance for a pair of optimal states~\cite{Wissmann2015a, Smirne2010a}, or as information flow from the bath back into the system~\cite{Chrui2018a, Lu2010a}. For the trace distance measure, by the concept that Markov dynamics is contractive for the trace distance in that the trace distance between any pair of initial states $\rho_1(0)$ and $\rho_2(0)$ do not increase under Markov dynamics, such that the non-Markovianity can be quantified by taking a maximum for all pairs of initial states and integrate over the region where the derivative of the trace distance is positive,
\begin{align}
	\max_{\rho_1, \rho_2}\int_{\delta_x>0} ds\delta_x(s, \rho_1, \rho_2),
\end{align}
where $\delta_x(t, \rho_1, \rho_2)=D'(\rho_1(t), \rho_2(t))$ is the derivative of the trace distance $D(\rho_1, \rho_2)=\mathrm{Tr}|\rho_1-\rho_2|/2$. For the TLS open dynamics considered here, the pair of optimal initial states that maximizes the integral can be chosen as~\cite{Ma2014g} $(|e \rangle \pm |g \rangle)/\sqrt{2}$. Moreover, the uncontrolled forward dynamics can be analytically solved. Tracing out the cavity degree of freedom, the density operator of the atom TLS $\rho_{a}$ can be written as (where we have taken $\omega = \omega_c=1$ without loss of generality) $\langle e | \rho_{a}(t) | e \rangle = |g(t)|^2 \langle e | \rho_{a}(0) | e \rangle $ and $\langle e | \rho_{a}(t) | g \rangle = g(t) e^{-i t} \langle e | \rho_{a}(0) | g \rangle $, with
\begin{align*}
	g(t) &= e^{-\lambda ^2 t / 4} \left[\frac{\lambda ^2 \sinh \frac{a t}{4}}{a}+\cosh \frac{a t}{4}\right], \quad % \nonumber \\
	a = \sqrt{\lambda ^4-16 \kappa^2},
\end{align*}
and the trace distance between the pair of optimal state is just $D(t)=|g(t)|$. In the orange shaded areas in Fig.~\ref{fig_petz_rev} (d) we show the regions where the trace distance increases, indicative of non-contractive dynamics.

The arise of non-Markovian dynamical features may be associated with the concept of an information backflow~\cite{Chrui2018a, Lu2010a, Zhong2013a} from the bath into the system. To quantify the information flow, here we use the quantum Fisher information~\cite{Petz2011a, Lu2010a, Zhong2013a}, which is often used to provide a lower bound for the mean-square error of some estimator of a parameter $\theta$ with the quantum Cram\'er-Rao bound~\cite{Holevo2011a,Helstrom1969a}. The quantum Fisher information for a parameter $\theta$ is defined as $\mathcal{F}_\theta = \mathrm{tr}\left[\rho(\theta) \bar{L}_\theta^2\right]$,
with the symmetric logarithmic derivative operator $\bar{L}_\theta$ given by $\partial_\theta \rho(\theta) = \{ \rho(\theta), \bar{L}_\theta \} /2$.
Here the parameter for the Fisher information is the angle $\theta$ in the initial state $|\psi(0) \rangle = \cos(\theta) |g0 \rangle + \sin(\theta)|e0 \rangle$. With the decomposition $\rho_{a}=\left( \mathbb{1} + \sum_{i=x,y,z} \vec{r}_i \sigma_i \right)/2$, $\vec{r}_i=\mathrm{tr}[ \sigma_i \rho_{a} ]$, the quantum Fisher information for TLS can be simplified~\cite{Zhong2013a, Liu2019a} as
\begin{align}
	\mathcal{F}_\theta = \left(\partial_\theta \vec{r}\right) \cdot \left(\partial_\theta \vec{r}\right) + \frac{(\vec{r} \cdot \partial_\theta \vec{r})^2}{1-|\vec{r}|^2}.
\end{align}
Substituting in the analytical solution of $\rho_a$, the Fisher information is then given by $\mathcal{F}_\theta = 4 |g(t)|^2$ for the uncontrolled open system dynamics. Since the trace distance between the two optimal initial states is given by $D(t) = |g(t)|$, it can be seen that they share the same monotonicity, showing that here the two measures of non-Markovianity is consistent: there would be a flow of information into the system measured by the Fisher information whenever the trace distance measure signifies non-contractive dynamics map.
We plot the dynamics of the quantum Fisher information for the $\theta$ parameter in Fig.~\ref{fig_petz_rev} (d), with $\theta=\pi/4$ for the numerical simulation. It can be observed that the Fisher information only increases in the non-contractive dynamics region, showing an agreement between these two interpretations of non-Markovianity. Here,  information generally flows from the system to the bath in the forward dissipative dynamics, and flows back to the system as a mirror image in the controlled reversal dynamics. Thus, by taking the TLS and the cavity as its immediate bath as a whole composite system with a given initial state, we may use the Petz map to recover the Markov dynamics of the composite system with~\eqref{eq_mmeqR}, where from the standpoint of the TLS, its non-Markovian dynamics is being reversed. It should also be pointed out that the reversal master equation~\eqref{eq_mmeqR} is constructed with a single initial state in mind, where the reference state for the Petz recovery map at each time step is based on the forward dynamics of this chosen initial state. Note that the Petz map may also be used with time-independent states~\cite{Kwon2022a, Lautenbacher2022a} as reference states to provide some (approximate) protection for states close to the reference states.
It is also possible to rotate out just the dissipative part and keep the ideal unitary evolution.
% In practical scenarios, one may often wish to not revert to the initial state, but just to cancel out the dissipative dynamics~\cite{Kwon2022a} and keep the desired unitary evolution of the system Hamiltonian $H$. One can rotate the reversal map such that the final state of the reversal map $\mathcal{L}'_B$ is not the initial state but the initial state evolving under the free closed system Hamiltonian $\mathcal{U}(\rho_0)$, where $\mathcal{U}=U(t) \rho_0 U^\dagger(t)$ and $U(t)=\mathcal{T}\exp \left[-i\int ds H_s\right]$.
To this end, one may decompose the rotated dynamical map as in~\cite{Kwon2022a}, which may also be constructed via a rotated density operator $\beta_{t'} = U(t') \rho^B_{t'} U^\dagger(t')$ where $U(t)=\mathcal{T}\exp \left[-i\int ds H_s(s)\right]$, such that $\beta(0)=\rho^B_0$ is the final state of the forward evolution $\rho_\tau$, and the final state $\beta(\tau)=U(\tau) \rho_{\tau}^B U^\dagger(\tau) = U(\tau) \rho_{0} U^\dagger(\tau)$ is the desired state: initial state evolving under the free $H$ for a duration of $\tau$.

\section{Conclusion}

The Jaynes-Cummings model is an important paradigm for both theoretical elegance and its viability of experimental implementations. Despite its seemingly simplicity, it is nevertheless able to provide rich physical phenomena. In this paper, we consider a scenario where the Jaynes-Cummings model is embedded in a non-Markovian bosonic bath, with a tunable memory effect parameter $\gamma$ that allows one to continuously tune the bath from one with strong memory effects (small $\gamma > 0$) all the way to a memory-less one ($\gamma \rightarrow \infty$). From the standpoint of the JC atom, it is experiencing a structured hierarchical environment: a direct one constituted of the JC model's cavity, and an indirect one constituted of the bosonic bath. This setup allows the JC atom to display memory effects even with a memory-less bath (and vice versa) with appropriate system parameters, and it also allows for exact analytical calculations.

In this paper, we considered two different ways to control and protect the quantum state of the JC atom in a new structured environment. The LEO approach focuses on the control applied on the system alone, and the control can be fault-tolerant. The effects of the control can be exactly calculated non-perturbatively, and with the help of partition maps and the QSD equation, we can analytically show the mechanism of the LEO is to suppress the dissipative terms in the master equation with an integral of a kernel over a fast oscillating function. When the parameters of the pulses are chosen to support this condition, the LEO can effectively decouple the pre-chosen projective space from leakage to the rest of the Hilbert space.
On the other hand, the Petz recovery map generally requires the ability to engineer system-bath coupling, in addition to controlling the system. It is then able to take the final state at time $\tau$ of the dissipative dynamics, and revert it back to the initial state, \textit{along the same path}: for the backwards propagation at time $t'$, $\rho^B_{t'} = \rho_{\tau - t'}$ of the forward evolving state. It is also possible to rotate the controls to only cancel out the dissipative part and keep the desired unitary evolution. The non-Markovian implications of the Petz recovery map is also considered, using both a trace-distance measure and one based on information flow between the system and bath. Another difference between the two approaches is that the Petz recovery map is usually applied at the end of a dissipative evolution, while the control of the LEO is generally being applied as the state is evolving. We found that both methods are effective at protecting the two-level atom from the environmental effects, and one may choose either method depending on the states to be protected and specific experimental realizations. It is worth noting while we consider a lossy cavity with initial vacuum cavity state for an analytical solution of the non-Markovian dynamics without losing features of the interesting physics, the decoherence control methods considered here are general, and can work for setups such as strong atom-cavity coupling ($\kappa/\omega > 1$) and initial states with non-zero photons in the cavity. While we focus on a lossy cavity open system model where the atom only directly interacts with the JC cavity mode~\cite{Carmichael2018a,Scala2007a}, the control may also be applied to systems with dissipative two-level systems in addition to lossy cavity, for example with a system-bath coupling like $L = a + \sigma_-$.
Finally, we want to emphasize that even after 60 years since its first publication, the Jaynes-Cummings (JC) model remains a vibrant area of research. Many recent theoretical and experimental works have shown that the JC model is still an active goldmine.

\section*{Acknowledgement}
This work is supported in part by ACC-New Jersey under Contract No. W15QKN-24-C-0004.

\section*{Disclosures}
The authors declare no conflicts of interest.

%%%%%%%%%%%%%%%%%%%%%%%%%%%%%%%%%%%%%

% \bibliographystyle{opticajnl}
% \bibliography{hjc_dctrl}

%%%%%%%%%%%%%%%%%%%%%%%%%%%%%%%%%%%%%

%%%%%%%%%%%%%%%%%%%%%%%%%%%%%%%%%%%%%

\end{document}